\documentclass[12pt]{article}
\usepackage{epsfig,graphicx,psfrag}
\usepackage{slashed} 
\usepackage{float}
\usepackage{bigdelim}
\usepackage{multirow}

\newcommand{\be}{\begin{equation}}
\newcommand{\ee}{\end{equation}}
\newcommand{\ba}{\begin{eqnarray}}
\newcommand{\ea}{\end{eqnarray}}

\newcommand{\MeV}{{\rm MeV}}
\newcommand{\GeV}{{\rm GeV}}
\newcommand{\ice}[1]{\relax}

\newcommand{\MSbar}{\overline{\rm MS}}

\usepackage{color}
\usepackage{amsmath}

\usepackage{booktabs} 
\begin{document}
\thispagestyle{empty}

\begin{flushright}
SI-HEP-2013-14 \\
QFET-2014-05
\end{flushright}
\phantom{}
\vspace{.2cm}

\begin{center}

{\Large\bf  
Radial excitations of heavy-light mesons \\from QCD sum rules }

\vspace{1cm}
{\bf
P.~Gelhausen, A.~Khodjamirian, A.A.~Pivovarov 
and D.~Rosenthal
}

\vspace{2mm}
{\it  
Theoretische Physik 1, 
Naturwissenschaftlich-Technische Fakult\"at,\\
Universit\"at Siegen, D-57068 Siegen, Germany }\\[2mm]


\begin{abstract}
QCD sum rules are commonly used to predict 
the characteristics of ground-state hadrons.
We demonstrate that two-point sum rules 
for the decay constants of charmed ($D^{(*)},D_s^{(*)}$) and bottom 
($B^{(*)},B_s^{(*)}$) mesons can also be modified to 
estimate the decay constants of the first radial excitations,
$D^{(*)'},D_s^{(*)'}$ and $B^{(*)'},B_s^{(*)'}$, respectively, provided 
the  masses of these resonances are used as an input.  
For the radially excited
charmed mesons we use available experimental data, whereas the masses
of analogous bottom mesons are estimated 
from the heavy-quark limit. The decay constants predicted 
for the radial excitations of heavy-light pseudoscalar and vector mesons 
are systematically smaller than those 
of the ground states 
and we comment on 
the possible origin of this difference.
Our results  
can be used in the sum rule calculations of heavy-to-light
form factors
and in the factorization approximations for nonleptonic
$B$-meson 
decays
where the decay constants of charmed mesons 
enter as input parameters.   
\end{abstract}

\end{center}

\section{Introduction}

The spectrum of hadrons with a given
spin-parity ($J^P$) and flavour
contains {\em radial excitations}, the sequential resonances heavier 
than the ground state.
In the $N_c\to \infty$ limit  of QCD a series of equidistant resonances 
is anticipated ~\cite{tHooft}.
Models of equidistant states 
based  on the resonance saturation of the two-point correlation functions 
are used to investigate quark-hadron duality and 
its violation~\cite{dualmod,Bigi:1998kc,Groote:2001im}.
Some recent applications to heavy-flavour decays can be found in 
\cite{BBNS,dualpeng}.  

A clear identification of radial excitations 
on the background of hadronic continuum is a difficult experimental task. 
Usually these resonances are strongly coupled to the two- and 
three-hadron states
from hadronic continuum, and these
couplings generate large total widths. 
Note also that a strong mixing 
via intermediate continuum states can in principle significantly 
influence the pattern of the excited resonances, affecting 
their masses, widths and decay constants. 
It is therefore not surprising that  
radial excitations are well established \cite{PDG} only 
for a few mesons.  
In this respect, the best studied are the neutral 
vector ($J^P=1^-$) mesons directly produced in $e^+e^-$ 
annihilation, especially
the heavy quarkonia. There are at least five (six) observed 
radial excitations\footnote{ 
~The notion ``radial excitations'' stems 
from quantum mechanics, where
it is used to denote the bound states obtained by
solving the radial Schr\"odinger equation in a central 
potential. In the framework of the quarkonium potential model,
the radially excited $2{}^3S_1$-state  differs from the orbitally 
excited ${}^3D_1$-state, both having  the same $J^P=1^-$. 
In QCD, a hadron is  a relativistic bound state with a certain 
valence quark  content. Hence,  here we count as radial excitations
all resonances having the same flavour and $J^P$ as
the ground state. }
of   the $J/\psi$  ($\Upsilon (1S)$) meson~\cite{PDG}. 
For the light-quark mesons some excited  states 
are presented in~\cite{PDG}, 
for example, $\rho(1450)$ and $\rho(1700)$ are identified as radial excitations of $\rho(770)$.

Relatively little is known about 
the radial excitations of heavy-light $D^{(*)}$ and $B^{(*)}$
mesons. On the experimental side, there are few observations of 
charmed resonances~\cite{PDG,BaBar,Bellerad,LHCb}
whose  quantum numbers and masses  fit the expected properties of 
the radially excited states  of $D$, $D^*$ and $D_s^*$, 
denoted here 
as $D'$, $D^{*'}$ and $D_s^{*'}$, respectively. 
These resonances decay strongly to 
ground states and light mesons with a width in the ballpark of 
100~MeV.
A pseudoscalar 
charmed-strange meson 
$D_s'$ was not yet established. 
Recent results of LHCb~\cite{LHCb} and 
CDF~\cite{Aaltonen:2013atp} collaborations
on the excited bottom mesons
hint at  $B'$ and $B^{*'}$ states.

Recently the radial excitations of charmed mesons 
were
discussed in connection with the semileptonic $B\to D'\ell \nu_\ell$ 
decays~\cite{Ligeti,Becirevic,Hernandez}, 
hence dynamical characteristics of these
mesons are becoming a topical subject for heavy flavour studies.  

The properties of radially excited heavy-light resonances were predicted in
various versions of the constituent  quark model, 
starting from~\cite{Godfrey};
for a recent analysis, see e.g.,~\cite{EFG10}. Assuming the
dominance of the valence 
quark-antiquark state and solving the bound-state problem 
for a certain quark-antiquark potential, one calculates
the energy spectrum and the values of the 
wave functions at the origin, i.e.,
the masses and decay
constants of radially excited mesons. 
However, the accuracy of the quark-model calculations is difficult 
to assess
in QCD, since  the relativistic quark-antiquark 
and gluon degrees of freedom beyond 
valence approximation are not explicitly included.
Some other theoretical studies were presented 
in~\cite{Badalian:2011tb,Sun:2014wea}.
Quite recently, lattice QCD studies 
were performed to establish 
the properties of radially excited resonances.
In particular, excited open-charmed mesons with $J^P=0^-,1^-$
have been studied 
in~\cite{Becirevic, Latt_Dpi, Latt_excited}.

In the QCD sum rule approach~\cite{SVZ},  a correlation function 
of two interpolating quark currents 
with a given flavour content and $J^P$ 
is calculated using the operator product 
expansion (OPE) in terms of QCD vacuum condensates. 
In the hadronic dispersion relation,  obtained 
applying unitarity to this correlation 
function, all radially excited states,
together with the continuum multi-hadron states with the quantum numbers 
of the  interpolating currents are usually included into one hadronic spectral density whereas the ground-state 
contribution is isolated.
The dispersion integral over the spectral density of excited 
and multi-hadron states 
is approximated applying the quark-hadron duality and 
introducing an effective threshold. 
After that the physical characteristics of the ground 
state are determined
with a certain accuracy. 

It is conceivable that a correlation function 
calculated in the spacelike region, via quark-hadron duality 
provides
certain dynamical information not only on the ground-state but on the 
whole spectrum of 
resonances\footnote{Attempts to describe
the properties of excited states 
using finite energy sum rules, that is, 
attributing a finite duality 
interval to each sequential resonance, were made already 
quite some time ago, see~\cite{FESR}.}. Indeed,
as shown in~\cite{dualmod}, 
a
correlation function, analytically continued to timelike momentum
transfers, yields an infinite  ``comb'' of 
equidistant poles.
On the other hand, it is evident that the OPE with truncated 
power corrections can only provide a  
very limited information about hadronic states. 
In particular, the strong couplings 
of resonances to continuum states, their mixing  and 
the resulting resonance widths are difficult to reproduce. 

In this paper we consider as a study case 
the QCD sum rules for heavy-light currents, interpolating 
 pseudoscalar and vector charmed and bottom mesons.  
The conjecture formulated above 
is  addressed only to the first radial excitations
of heavy-light mesons.  We demonstrate that modifying 
QCD sum rules, it is possible to determine the decay  constants 
of these states in addition to the ones of ground states.
In what follows, we extensively use the results of 
our recent work~\cite{GKPR}
where the sum rules for pseudoscalar 
and vector heavy-light mesons were updated.   

The  plan of the paper is as follows. In Section~2 we outline
various procedures of constraining and/or estimating the decay 
constants of the first radially excited  states
using the sum rules. In Section~3 we present the  
numerical analysis. In Section~4 we summarize the 
results and discuss their possible applications.

\section{Including  radial excitations in the sum rules}

In what follows, we consider  QCD sum rules for the
heavy-light pseudoscalar $(J^P\!=\!0^-)$ 
and vector ($J^P=1^-$) mesons, 
which are, respectively, obtained 
from the following correlation functions:
\be
\Pi_5(q)=i\int d^4 x \, e^{iqx} \langle 0| 
T \{ j_5(x) j^\dagger_5(0) \} |0 \rangle \,
\label{eq:corr5}
\ee
and 
\begin{align}
\Pi_{\mu\nu}(q)=i\int d^4x\, e^{iqx} \langle 0| 
T \{j_\mu(x) j^\dagger_\nu(0)\}|0 \rangle=
\Big(\!- g_{\mu\nu}q^2+q_\mu q_\nu\Big)\widetilde{\Pi}_T(q^2)+
q_\mu q_\nu\Pi_L(q^2)\, ,
\label{eq:corrV}
\end{align}
where  $j_5=(m_Q+m_q)\bar{q}\,i\gamma_5 Q $ and 
$j_{\mu }=\bar{q}\gamma_\mu Q $ are 
the interpolating quark currents,  $Q=c,b$ and  $q=u,d,s$
are the quark fields with finite quark masses defined 
in $\MSbar$-scheme.
In~(\ref{eq:corrV}), 
only $J^P=1^-$ states contribute to the invariant amplitude 
multiplying the transverse
kinematic structure,
and we define $\Pi_T(q^2)\equiv q^2\widetilde{\Pi}_T(q^2)$.
The decay constants of ground-state mesons $H=\{ B,D\} $  
and $H^*=\{B^*,D^*\}$ 
are defined in a standard way, 
\be
\langle 0|j_5|H(q) \rangle = m_{H}^2 f_{H},\quad 
\langle 0|j_\mu|H^*(q) \rangle = m_{H^*}\epsilon^{(H^*)}_\mu f_{H^*} , 
\label{eq:fBst}
\ee
where $\epsilon^{(H^*)}_\mu $ is the polarization vector of $H^*$.
The same definitions are used for the decay constants 
$f_{H'}$ and $f_{H^{*'}}$ of the 
radially excited states $H'$ and $H^{*'}$ with the masses
$m_{H'}$ and $m_{H^{*'}}$, respectively. 

The correlation functions are calculated at $q^2\ll m_Q^2$, 
using OPE which contains
the perturbative part  with $\mathcal{O}(\alpha_s^2)$ (NNLO) 
accuracy
and   the vacuum condensate contributions up to dimension $d=6$:
\be
\Pi^{OPE}_{T(5)}(q^2)=\Pi_{T(5)}^{(pert)}(q^2)+
\Pi_{T(5)}^{\langle \bar{q} q \rangle} (q^2)+
\Pi_{T(5)}^{\langle d456\rangle} (q^2)\,.
\label{eq:OPE}
\ee
In the above, the quark condensate contribution (including the NLO terms calculated in \cite{GKPR})  and 
the sum of gluon, quark-gluon and four-quark condensate contributions 
with dimensions $d=4,5,6$ have, respectively, the indices 
$ \langle \bar{q} q \rangle $ and $\langle d456 \rangle$.
The expressions entering the OPE (\ref{eq:OPE}) can be found in~\cite{GKPR}
and we will not repeat them here for brevity.

Turning to the hadronic representations of the correlation functions, 
we  modify them, explicitly separating the first radial excitation 
from the spectrum  of heavy-light states.
The hadronic spectral densities of the correlation functions 
receive then the following form:
\ba
\rho_5(s)\equiv \frac{1}{\pi}\mbox{Im} \Pi_{5}(s)=
m_H^4 f_{H}^2\delta(s-m_H^2)+ 
m_{H'}^4f_{H'}^2\frac{\Gamma_{H'}m_{H'}}{\pi[(m_{H'}^2-s)^2
+\Gamma_{H'}^2m_{H'}^2]}
\nonumber\\
+ 
\tilde{\rho}{\,}^h_5(s)\theta(s-(m_{H^*}+m_P)^2)\,,
\label{eq:rhoH}
\ea 
and 
\ba 
\rho_{T} (s)\equiv \frac{1}{\pi}\mbox{Im} \Pi_{T}(s)=
m_{H^*}^2 f_{H^*}^2\delta(s-m_{H^*}^2) 
+m_{H^{*'}}^2 f_{H^{*'}}^2
\frac{\Gamma_{H^{*'}} m_{H^{*'}}}{\pi[(m_{H^{*'}}^2-s)^2
+\Gamma_{H^{*'}}^2m_{H^{*'}}^2]}
\nonumber
\\
+ \tilde{\rho}{\,}^h_T(s)\theta(s-(m_H+m_P)^2)\,,
\label{eq:rhoHstar}
\ea
where $P$ is the light pseudoscalar meson (pion or kaon).
For the excited resonances we assume a Breit-Wigner (BW) 
form of the spectral density 
with a constant total width. The dependence on the width 
and the modification of the BW form 
with an energy-dependent width will also be investigated. 
The spectral densities $\tilde{\rho}{\,}^h_{5,T}(s)$ include
the contributions of the excited states 
located above the first radial excitation and the continuum states.
The latter start from the two-hadron thresholds  $s=(m_{H^*}+m_P)^2$  
($s=(m_H +m_P)^2$) in the pseudoscalar (vector) channel. For pseudoscalar channels we take
the decay $H \rightarrow H^* P$ as a physical process giving the threshold.
For the vector channel we chose the process $H^* \rightarrow H P$.
Note that the widths of the excited resonances $H^{(*)'}$ are 
generated by their strong couplings to the continuum states,
hence a part of the continuum contribution is 
effectively included in the radially excited 
resonance terms in the above 
spectral densities. 

Our main assumption is that the semilocal quark-hadron 
duality approximation remains valid
after isolating the excited state from the hadronic sum:
\ba
\tilde{\rho}^h_{5}(s)\theta(s-(m_{H^{*}}+m_P)^2)= 
\rho^{(pert)}_{5} (s) \theta(s-\tilde{s}^{H}_0)\,,\\
\label{eq:dual5}
\widetilde{\rho}^h_{T}(s)\theta(s-(m_{H}+m_P)^2)= 
\rho^{(pert)}_{T} (s) \theta(s-\tilde{s}^{H^{*}}_0)\,,
\label{eq:dualT}
\ea
where $\rho^{(pert)}_{5,T}(s)=(1/\pi) \mbox{ Im } \Pi^{(pert)}_{5,T}(s)$ 
is the spectral density 
of the perturbative loop contributions to the OPE and 
$\widetilde{s}^{H^{(*)}}_0$
is  the effective threshold. The latter parameter is expected 
to be larger than the one
used in the conventional sum rules where only the ground state
is separated from the hadronic sum.
After applying 
the above duality ansatz and Borel transformation,
the resulting sum rules are:
\begin{align}
\label{BorelSRP}
f_{H}^2 m_{H}^4e^{-\frac{m_{H}^2}{M^2}}+
f_{H^{'}}^2m_{H^{'}}^4
\!\!\!\!\!\!\!\int\limits^\infty_{(m_{H^*}+m_P)^2} 
\!\!\!\!\!\!\!\!ds\; 
e^{-\frac{s}{M^2}} \frac{m_{H^{'}}
  \Gamma_{H'}}{\pi[(s-m_{H^{'}}^2)^2+m_{H^{'}}^2 
\Gamma_{H'}^2]}\nonumber\\
=\widetilde\Pi^{(pert)}_5(M^2,\widetilde{s}_0^{H})+\widetilde\Pi_{5}^{\langle \bar{q} q
  \rangle} 
(M^2)+
\widetilde\Pi_{5}^{\langle d456\rangle} (M^2)\,,
\end{align}
\begin{align}
\label{BorelSRV}
f_{H^*}^2 m_{H^*}^2e^{-\frac{m_{H^*}^2}{M^2}}+
f_{H^{*'}}^2m_{H^{*'}}^2\!\!\!\!\!\!\!\int\limits^\infty_{(m_H+m_P)^2} 
\!\!\!\!\!\!\!ds\; e^{-\frac{s}{M^2}} 
\frac{m_{H^{*'}} \Gamma_{H^{*'}}}{\pi[(s-m_{H^{*'}}^2)^2+m_{H^{*'}}^2 
\Gamma^2_{H^{*'}}]}
\nonumber\\
=\widetilde\Pi^{(pert)}_T(M^2,\widetilde{s}_0^{H^*})+\widetilde\Pi_{T}^{\langle \bar{q} q \rangle} (M^2)+
\widetilde\Pi_{T}^{\langle d456\rangle} (M^2)\,,
\end{align}
where the shorthand notation
\be
\widetilde\Pi_{T(5)}^{(pert)}(M^2,s_0)=\!\!\!\!\!\!\!\int\limits_{(m_Q+m_q)^2}^{s_0}
\!\!\!\!\!\!\!ds \, e^{-s/M^2}\rho^{(pert)}_{T(5)} (s) 
\label{eq:Pipert}
\ee
 is used and the Borel transformed correlation function is denoted as $\widetilde\Pi_{T(5)}(M^2)$.

The masses and total widths of excited mesons
$H'$ and $H^{*'}$ will be specified in the next 
section using experimental data on 
charmed states and heavy-quark symmetry relations. 
As usual in the sum rule analysis, one has to adopt an 
optimal interval of 
the Borel parameter values 
$\Delta M^2\equiv \{ M^2_{min}\div   M^2_{max}\}$,
where the lower (upper) boundary is chosen so that within this interval 
the OPE is reliable (the continuum contribution remains subleading).
After that the above sum rules  can be used  to 
estimate the meson decay constants.

As a starting point  we fix the decay constants of the ground states. 
We use the decay constants $f_{H}$ and $f_{H^*}$ obtained 
in~\cite{GKPR} from 
conventional QCD sum  rules (where the first excitation 
is included in the duality ansatz). 
Note that  within uncertainties these values  
are in agreement  with more accurate recent
results from lattice QCD and also with the decay constants of charmed mesons 
extracted from experiment. Hence, lattice and/or experimental input values 
for the ground-state decay constants can equally well be used.
After fixing the input values for the masses and ground state residue,
the two unknown parameters remain in each sum 
rule~(\ref{BorelSRP}) or~(\ref{BorelSRV}):
the decay constant  $f_{H^{(*)'}}$  of the excited state we are interested in 
and the new effective threshold $\tilde{s}^{H^{(*)}}_0$. 
Putting the latter
threshold to infinity, one immediately obtains 
the  upper bound for $f_{H'}$ or $f_{H^{*'}}$.
This bound is based on 
the positivity of all contributions to the hadronic spectral density 
and is independent of the duality approximation.

To employ the sum rules~(\ref{BorelSRP}) and~(\ref{BorelSRV}),
while fixing the ground-state decay constants, one could try to follow 
the standard procedure (see e.g.,~\cite{GKPR}): 
adjusting the effective threshold 
to the mass of the excited state. The mass
squared of the resonance can be calculated 
by dividing   the sum rule 
differentiated over $(-1/M^2)$ by the original sum rule.
However, this procedure demands a very accurate knowledge 
of the excited meson mass, and practically only works
in the zero width approximation when the spectral density of the excited state
reduces to a delta function. Hence, in our numerical analysis presented below we use two 
different procedures:

(I)  The decay constants $f_{H^{(*)}}$ and $f_{H^{(*)'}}$ 
and the effective threshold 
 $\tilde{s}^{H^{(*)}}_0$ are fitted simultaneously. 
To this end for each separate heavy-light channel, 
we minimize the squared difference between the l.h.s. and r.h.s. of the sum rule 
summed over several points within $\Delta M^2$. E.g., for the 
pseudoscalar heavy-light meson channel we fit the values 
$f_{H},f_{H'}$ and $\widetilde{s}_0^{H}$ from
\be
  \sum_i \Big| f_{H}^2 m_{H}^4 e^{-\frac{m_{H}^2}{M_i^2}} +
f_{H'}^2 m_{H'}^4 e^{-\frac{m_{H'}^2}{M_i^2}} 
-\Big[\widetilde\Pi_5^{(pert)}(M_i^2, \tilde{s}^H_0) +\widetilde\Pi_{5}^{\langle \bar{q} q \rangle} (M_i^2)+
\widetilde\Pi_{5}^{\langle d456\rangle} (M_i^2)\Big]\Big|^2 = min\,,
\label{eq:min}
\ee
%
where for brevity the width of the excited state is neglected
being included in the 
numerical  analysis. A similar minimization procedure
is used for the vector meson channel.

(II)  The ground state  contribution is eliminated 
from the sum rule. This is done,
multiplying the correlation function by 
the factor $(m_{H^{(*)}}^2-q^2)$ before  applying 
the Borel transformation. An equivalent procedure
is to act with the differential operator 
$[\frac{d}{d(1/M^2)}+m_{H^{(*)}}^2]$   on the sum 
rules~(\ref{BorelSRP}) and~(\ref{BorelSRV}). The correlation function is accordingly
modified, so that the perturbative part~(\ref{eq:Pipert}) 
contains an additional factor $(m_{H^{(*)}}^2-s)$ under the 
integration.
The resulting sum rule relations are then used
to fit the decay constant of the excited state and the effective threshold
with the  minimization  similar to~(\ref{eq:min}) where now only the 
excited state contribution is present. 

We emphasize that the method used in this paper goes beyond 
the conventional sum rule technique. E.g., in the procedure (I) described above, 
the two decay constants and effective threshold are simultaneously fitted 
to the Borel-transformed correlation function. 
Hence, cautiously, one cannot exclude that a ``systematic'' uncertainty related to 
the quark-hadron duality is larger than in the usual 
sum rules. Still, due to the positivity of the hadronic spectral function, a
cancellation between ground- and excited-state contributions in the sum rule 
cannot take place, hence a significantly biased estimate of decay constants is excluded. 
Furthermore, an important indication of the reliability is provided if both procedures (I) and (II)  
reproduce reasonably close values for the decay constant of a radially excited state.

\section{Numerical estimates}
 
The input parameters in the OPE  on the r.h.s. of 
the sum rules~(\ref{BorelSRP}) and~(\ref{BorelSRV}) 
include the quark masses, 
strong coupling and condensate densities. We adopt the same values as in 
Table~I of~\cite{GKPR}
where one can find the detailed discussion and relevant references. 
In particular,  we use the $\overline{\rm MS}$ 
values of the quark masses:
$\overline{m}_b(\overline{m}_b)=4.18\pm 0.03\,$~GeV, 
$\overline{m}_c(\overline{m}_c)=1.275\pm0.025\,$~GeV,
$\overline{m}_s(2\,\GeV)=  95 \pm 10 $~MeV, the strong coupling
$\alpha_s(M_Z)=0.1184\pm 0.0007$,
and the 
quark condensate density 
$\langle\bar{q}q\rangle(2 \GeV) =-(277^{+12}_{-10}\,\text{MeV})^3 $. 
We adopt  the same default renormalization scale $\mu=1.5$~GeV  
($\mu
= 3$ GeV)   for the $c$-quark ($b$-quark) correlation function 
as in~\cite{GKPR}
allowing further to vary it within the intervals 
$1.3~\GeV \div 3~\GeV$  
($3~\GeV \div 5~\GeV$). 
The interval of the Borel parameter squared 
used in the sum rules for charmed mesons is 
$M^2=2.5\div 3.5$~GeV$^2$.
We also vary it to $M^2=2.0\div 3.0$~GeV$^2$ and 
$M^2=3\div 4$~GeV$^2$ in order to estimate the related 
uncertainty of the results. For the bottom
mesons 
we use $M^2=6.0\div 8.0$~GeV$^2$ 
as a default interval, shifting it to 
$M^2=5.5\div 7.5$~GeV$^2$ and $M^2=6.5\div 8.5$~GeV$^2$ for 
uncertainty estimates.
The default Borel intervals still satisfy the criteria mentioned in the previous
section  although they are shifted to somewhat larger values
with respect to the intervals used 
in~\cite{GKPR}. This is done 
on purpose in order to enhance the contributions of 
the excited states.

In the  hadronic part of 
the sum rule the masses and total widths of the three 
excited charm mesons are 
taken from experiment~\cite{PDG} 
and collected in Table~\ref{tab:masses}. In particular, 
let us mention that the BaBar collaboration~\cite{BaBar} 
observed two candidates for the
radially excited charmed mesons:
$D(2550)$ with $J^{PC}=0^-$ and $D^*(2600)$ with $J^{PC}=1^-$. 
The charmed-strange radially excited state  ${D^{*}_s}'(2700)$
with $J^P=1^-$  
was observed by several 
experiments~\cite{PDG,Bellerad,LHCb}.
To specify the mass of the remaining  
charm-strange pseudoscalar meson 
we use an estimate 
\be
m_{D_s^{'}}-m_{D_s} \approx m_{D^{'}}-m_{D}\, ,
\ee
relying on the $SU(3)_{fl}$ symmetry for the excitation 
energy.
For the radially excited $B$-mesons 
it is conceivable to apply simple relations 
valid in the heavy quark limit:
\be
m_{B^{(*)'}_{(s)}}-m_{B_{(s)}^{(*)}} \approx 
m_{D_{(s)}^{(*)'}}-m_{D_{(s)}^{(*)}}\, .
\label{eq:masses}
\ee
\begin{table}[h!]
\centering
\begin{tabular}{|c|ccc|c|c|c|} \hline
 \rule[-1mm]{0mm}{5mm} $H^{(*)'}$ & $m_{H'}$ [MeV] &
 $\Gamma_{H^{(*)'}}$ [MeV]  
& &  $m_{H^{(*)'}}-m_H$ [MeV] & $m^{QM}_{H^{(*)'}}$  [MeV]  \\\hline\hline
 \rule{0mm}{5mm} $D'$  & 2539~$\pm$~8
& 130  $\pm$  18 & \cite{PDG,BaBar} &669~$\pm$~8& 2581  \\
\rule{0mm}{5mm} $D^{*'}$  & 2612 $\pm$ 6 & 93 $\pm$ 14 &
\cite{PDG,BaBar}   
&601~$\pm$~6&2632 \\
\rule{0mm}{5mm} $D'_s$  & \underline{2618~$\pm$~50} 
& \underline{100 $\pm$ 50} &  & \underline{650~$\pm$~ 50}& 2688 \\
\rule{0mm}{5mm} $D_s^{*'}$ &2709~$\pm$~4& 117~$\pm$~13
& \cite{PDG} &597~$\pm$~4 & 2731  \\
\hline
\rule{0mm}{5mm} $B'$ & \underline{5929 $\pm$  50} &
\rdelim\}{4}{2.2cm}[\underline{$100\pm50$}]        &
&\rdelim\}{4}{2.2cm}[\underline{$650\pm50$}]   
& 5890\\
 \rule{0mm}{5mm} $B^{*'}$ & \underline{5975 $\pm$  50} & &  &  & 5906 \\
\rule{0mm}{5mm} $B'_s$ & \underline{6017 $\pm$  50}  & & & &5976  \\
\rule{0mm}{5mm} $B_s^{*'}$ &\underline{6065 $\pm$  50}  &  &  &  & 5992  \\
 \hline\hline
\end{tabular}
\caption{\it Masses and total widths  of  the first radially excited   heavy-light
mesons and the mass shifts with respect  to the ground-state mesons. The
underlined masses and widths are our estimates for the yet unobserved
resonances. The quark model predictions \cite{EFG10} are shown in the last
column.}
\label{tab:masses}
\end{table}

The estimated masses of excited hadrons are shown 
underlined in Table~\ref{tab:masses} and compared to 
the quark model estimates from ref.~\cite{EFG10}.
Note that  quark model predictions are for excited charmed mesons
systematically larger  than the available experimental values  and for
bottom mesons systematically smaller than the masses
estimated from  heavy-quark symmetry relations.
The accuracy of symmetry
relations for $B$-mesons is expected to be higher than for 
charmed mesons since
the corrections are of $\mathcal{O}(1/m_Q)$. 
The recently observed state $B(5970)$~\cite{Aaltonen:2013atp}
interpreted as an excitation with  $J^P=1^-$, nicely coincides 
with
${B^*}'$-meson predicted from~(\ref{eq:masses})
with an 
estimated mass $m_{{B^*}'}=5975~\MeV$. 
Hence, the relations
based on heavy-quark symmetry seem to be reliable.
We add $\pm 50~\MeV$ 
uncertainty to the central
values of the estimates~(\ref{eq:masses}) to allow for 
a very conservative error. 
Furthermore, for the yet unknown total widths 
we assume a rather broad interval 
between 50~MeV and 150~MeV 
which is in the ballpark of measured
total widths of charmed mesons.
Note that an accurate prediction of the total widths of 
radially excited
states is a difficult task because several channels of strong 
(flavour-conserving) 
decays with the corresponding strong couplings
contribute.
In future, when the masses, branching fractions 
and total widths of all radially 
excited heavy-light 
mesons  will be measured, one 
can substantially refine the above input.  

The results for the decay constants for heavy-light mesons 
predicted from  QCD sum rules~(\ref{BorelSRP}) and~(\ref{BorelSRV})
are collected in Table~\ref{tab:bottom}.
In the first column we quote the decay constants of ground-state
mesons obtained in~\cite{GKPR}.
We use these constants as inputs  while obtaining the upper bounds 
for the decay constants of excited states which 
are independent of duality assumption. These bounds are calculated 
putting $\widetilde{s}_0^{H^{(*)}} \to \infty$ in the sum 
rules~(\ref{BorelSRP}) and~(\ref{BorelSRV}). 
For each bound the maximal value is determined within the optimal Borel
interval.
To this value we also add the uncertainty obtained after varying the
parameters in the sum rules. The resulting bounds are presented in the last column of Table~\ref{tab:bottom}.
We see that the bounds are somewhat
restrictive for the charmed mesons, but not for bottom mesons.

Our main numerical results obtained from the 
fit procedures (I) and (II) described in 
the previous section are also shown in Table~\ref{tab:bottom}. 
Remarkably, these two quite different   
procedures predict consistent values of the decay constants 
of excited mesons. 
Most importantly, the ground-state decay constants 
obtained from the fit (I) are very close to the ones 
obtained in~\cite{GKPR}
from conventional sum rules, providing a cross-check 
of our calculation and 
ensuring the validity of quark-hadron 
duality beyond the ground state. 

The uncertainties quoted in Table~\ref{tab:bottom}
originate from:
a) the variation of all input parameters in the OPE; 
b) the shift of the $M^2$ intervals as explained above;
c) the mean squared fit error (reflecting the uncertainty induced by duality threshold);
d) the variation of the masses of excited states and widths
within the intervals shown in Table~\ref{tab:masses}. 
Note that we prefer a rather conservative estimate and do not account for correlations 
between separate  uncertainties  adding them all in quadrature.
\begin{table}[h!]
\centering
\begin{tabular}{|c||c||c|c|c||c|c||c|} \hline
\rule[-1mm]{0mm}{5mm} \multirow{3}{*}{Meson}  &ref. \cite{GKPR}&\multicolumn{3}{c||}{Procedure (I)}&\multicolumn{2}{c||}{Procedure (II)}  & Upper bound\\
\cline{2-8}
 \rule[-1mm]{0mm}{5mm} & $f_{H^{(*)}}$ & $f_{H^{(*)}}$ & $f_{H^{(*)'}}$ 
&  $\tilde{s}^{H^{(*)}}_0$ & $f_{H^{(*)'}}$ & $\tilde{s}^{H^{(*)}}_0$ & $f_{H^{(*)'}}$ 
\\
 \rule[0mm]{0mm}{5mm}& $[\text{MeV}]$ & $ 
[\text{MeV}]$ & $ 
[\text{MeV}]$ & $ 
[\text{GeV}^2]$ & $ 
[\text{MeV}]$ 
&  $ 
[\text{GeV}^2]$ & $ 
[\text{MeV}]$ \\\hline\hline
 \rule{0mm}{5mm} $D^{(')}$ & 201$^{+12}_{-13}$  & 194$^{+6}_{-6}$ &
 137$^{+10}_{-23}$ 
& 7.24 & 138$^{+10}_{-22}$ & 7.24 & 189 \\\hline
 \rule{0mm}{5mm} $D_s^{(')}$ & 238$^{+13}_{-23}$ & 230$^{+7}_{-9}$ &
 143$^{+19}_{-31}$ 
& 7.48 & 146$^{+12}_{-36}$ & 7.49 & 219 \\\hline
 \rule{0mm}{5mm} $D^{*(')}$ & 242$^{+20}_{-12}$  & 235$^{+25}_{-12}$  
& 182$^{+12}_{-27}$ & 7.43 & 183$^{+13}_{-24}$ & 7.44 & 275 \\\hline
 \rule{0mm}{5mm} $D^{*(')}_s$ & 293$^{+19}_{-14}$ & 279$^{+21}_{-12}$ 
& 174$^{+22}_{-45}$ & 7.87 & 178$^{+20}_{-39}$ & 7.88 & 265 \\
 \hline\hline
 \rule{0mm}{5mm} $B^{(')}$ & 207$^{+17}_{-9}$  & 200$^{+18}_{-10}$ &
 163$^{+10}_{-11}$ 
& 36.75 & 166$^{+9}_{-10}$ & 36.78 & 279 \\\hline
 \rule{0mm}{5mm} $B_s^{(')}$ & 242$^{+17}_{-12}$ & 234$^{+15}_{-11}$ &
 174$^{+19}_{-19}$ 
& 37.72 & 178$^{+19}_{-17}$ & 37.75 & 320 \\\hline
 \rule{0mm}{5mm} $B^{*(')}$ & 210$^{+10}_{-12}$  & 208$^{+12}_{-21}$  &
 163$^{+54}_{-13}$ 
& 36.70 & 165$^{+46}_{-12}$ & 36.71 & 314 \\\hline
 \rule{0mm}{5mm} $B^{*(')}_s$ & 251$^{+14}_{-16}$ & 244$^{+13}_{-26}$ &
 190$^{+67}_{-20}$ 
& 38.58 & $194^{+57}_{-18}$ & 38.61 & 325 \\
 \hline\hline
\end{tabular}
\caption{\it 
Decay constants of charmed and bottom mesons
obtained from QCD sum rules and the corresponding 
effective thresholds.}
\label{tab:bottom} 
\end{table}

The estimated uncertainties 
for the ground-state decay constants obtained here and 
in~\cite{GKPR} are in the same ballpark.
For radially excited states the uncertainties of decay constants
are intuitively  expected to be larger than for the ground states.
In fact, the contributions of excited resonances to the Borel-transformed correlation
function have a smaller but comparable exponential weight with respect to 
the ground state, therefore 
the uncertainty returned by fit procedures (I) and (II)  can also be in the same ballpark,
as for example in the case of $B'$ meson. The actual outcome
of our numerical analysis becomes more transparent in terms of relative uncertainties, 
that is, if one divides the total variation quoted in Table 2 by the value obtained at the 
central input. These relative uncertainties  
vary from about 12 \% for the $B'$-meson up to 40\%  for the $D_s^{*'}$ and $B_{(s)}^{*'}$ 
mesons. Meanwhile, the corresponding uncertainties for all heavy-light ground states do  not exceed 15\%.  Note that the  ``systematic'' uncertainty  
caused by quark-hadron duality approximation, which is difficult to quantify 
on the basis of the input parameter variations, can also be somewhat larger 
for the decay constants of  the excited states.  

In Table~\ref{tab:bottom-err}
we present the error budget of our predictions in more detail.
The largest uncertainties originate from the renormalization scale and quark 
mass variation, added in quadrature together with other input parameters in the OPE and 
denoted as $\Delta_{param}$ in Table~\ref{tab:bottom-err}. 
The variations of decay constants introduced
by  the choice of the Borel window ($\Delta_{M^2}$), fit procedure ($\Delta_{fit}$), 
uncertain masses ($\Delta_{m_{H'}}$) and widths ($\Delta_{\Gamma}$)
of the excited resonances are smaller.
\begin{table}[t]
\centering
{\large
\begin{tabular}{|c||c||c|c||c|c|} \hline
 \rule[-1mm]{0mm}{5mm} Meson & $\Delta_{\rm param}$
 & $\Delta_{M^2}$ & $\Delta_{\rm
  Fit}$ & $\Delta_{\Gamma}$
& $\Delta_{m_{H'}}$\\\hline\hline
 \rule{0mm}{5mm} $D'$ & $^{+9}_{-21}$ \big($^{+9}_{-20}$\big)  
& $^{+5}_{-10}$ \big($^{+5}_{-8}$\big) & $^{+1}_{-1}$
\big($^{+1}_{-1}$\big) 
& $^{+1}_{-1}$ \big($^{+1}_{-1}$\big) & {\small{$<\pm1$ ($<\pm1$)}} \\\hline
 \rule{0mm}{5mm} $D_s'$ & $^{+12}_{-27}$ \big($^{+7}_{-29}$\big)
& $^{+7}_{-14}$ \big($^{+6}_{-16}$\big)& $^{+1}_{-1}$
\big($^{+2}_{-2}$\big) 
& $^{+8}_{-8}$ \big($^{+4}_{-14}$\big)& $^{+9}_{-4}$ \big($^{+7}_{-4}$\big)\\\hline
 \rule{0mm}{5mm} $D^{*'}$ & $^{+11}_{-24}$ \big($^{+10}_{-22}$\big)  
& $^{+8}_{-12}$ \big($^{+7}_{-10}$\big)  & $^{+2}_{-2}$
\big($^{+3}_{-3}$\big) 
& $^{+1}_{-1}$ \big($^{+1}_{-1}$\big) &  {\small{$<\pm1$ ($<\pm1$)}} \\\hline
 \rule{0mm}{5mm} $D^{*'}_s$ & $^{+14}_{-34}$ \big($^{+13}_{-31}$\big) 
& $^{+15}_{-29}$ \big($^{+13}_{-23}$\big) & $^{+5}_{-5}$
\big($^{+7}_{-7}$\big) 
& $^{+2}_{-2}$ \big($^{+2}_{-2}$\big) &  {\small{$<\pm1$ ($<\pm1$)}} \\
 \hline\hline
 \rule{0mm}{5mm} $B'$ & $^{+7}_{-8}$ \big($^{+7}_{-7}$\big)  & $^{+5}_{-7}$ \big($^{+5}_{-6}$\big) & $^{+1}_{-1}$ \big($^{+2}_{-2}$\big) & $^{+4}_{-4}$ \big($^{+4}_{-4}$\big) & $^{+2}_{-1}$ \big($^{+1}_{-  1}$\big) \\\hline
 \rule{0mm}{5mm} $B_s'$ & $^{+11}_{-11}$ \big($^{+9}_{-10}$\big)
& $^{+8}_{-10}$ \big($^{+7}_{-8}$\big)& $^{+2}_{-2}$
\big($^{+3}_{-3}$\big) 
& $^{+11}_{-10}$ \big($^{+11}_{-10}$\big)
& $^{+9}_{-1}$ \big($^{+10}_{-1}$\big)\\\hline
 \rule{0mm}{5mm} $B^{*'}$ & $^{+53}_{-12}$ \big($^{+46}_{-10}$\big)  & $^{+4}_{-4}$ \big($^{+3}_{-3}$\big)  & $^{+1}_{-1}$ \big($^{+2}_{-2}$\big) & $^{+4}_{-4}$ \big($^{+4}_{-4}$\big) & $^{+5}_{-3}$ \big($^{+4}_{-3}$\big) \\\hline
 \rule{0mm}{5mm} $B^{*'}_s$ & $^{+66}_{-15}$ \big($^{+56}_{-13}$\big) & $^{+8}_{-11}$ \big($^{+7}_{-9}$\big) & $^{+2}_{-2}$ \big($^{+3}_{-3}$\big) & $^{+7}_{-7}$ \big($^{+7}_{-7}$\big) & $^{+4}_{-1}$ \big($^{+3}_{-1}$\big) \\
 \hline\hline
\end{tabular}
}
\caption{\it Separate uncertainties for 
decay constants of  heavy-light excited  mesons from 
the sum rules applying the procedure (I) ((II)). All numbers are in MeV.}
\label{tab:bottom-err} 
\end{table}

As an alternative to a constant total width for excited heavy-light mesons, 
one can also use an energy-dependent width
taking into account the hadronic continuum threshold.
To investigate the influence of this effect we have inserted 
$\sqrt{s}\Gamma_{H^{(*)'}}(s)$ instead of  
$m_{H^{(*)}}\Gamma_{H^{(*)'}}$ in the Breit-Wigner ansatz for spectral
densities~(\ref{eq:rhoH}) and~(\ref{eq:rhoHstar})
where for the  $s$-dependent width the model~\cite{Kuhn:1990fe}  has been adopted
(see, e.g., also~\cite{BruchKK}), e.g., for $H^{*'}$: 
\begin{align}
 \Gamma_{H^{*'}}(s)= \Gamma_{H^{*'}}\frac{m_{H^{*'}}^2}{s} 
\left(
\frac{\lambda(s, m^2_H, m_P^2)}
{\lambda(m_{H^{*'}}^2, m_H^2, m_P^2)}\right)^{\frac{3}{2}}\;.
\label{eq:width}
\end{align}
In the above, $\lambda(a,b,c)=a^2+b^2+c^2-2ab-2bc-2ac$ and the kinematical factor
originates from the $p$-wave phase space of the decay $H^{*'}\to H P$. 
The analogous formula 
is valid for the excited pseudoscalar mesons where the width 
is dominated by 
the $H'\to H^* P$ decay. Altogether, the influence of 
the total width on the decay 
constants of excited mesons obtained from the sum rules is small.
To give an example,  the $D_s^{'}$-meson  decay constant obtained form the fit procedure (I) for the central input is 
 $f_{D_s^{'}} =  143 \pm 8$~MeV, where here the uncertainty corresponds to
varying only the total width within the interval $100\pm 50$~MeV.
This value  
shifts to $f_{D_s^{'}}=  148\pm16$~MeV  if the energy-dependent 
width~(\ref{eq:width}) is used in the fit. 
Note that neglecting  the total width of $D_s^{'}$ altogether
 yields $f_{D_s^{'}}= 128 $~MeV, a relatively small change.

Importantly, as seen from Table \ref{tab:bottom}, the decay constant of a heavy-light 
excited meson is predicted systematically  smaller
than for its ground state. 
To investigate the origin of this difference we present in Table~\ref{tab:rel}
the results of the numerical procedure (I) 
where separate contributions to OPE
are switched off. We notice that when only 
the perturbative 
contributions are left and condensate contributions are neglected, 
the 
decay constants become numerically closer. 
\begin{table}[h!]
\begin{center}
\begin{tabular}{|c|c|c|} \hline
\rule[-1mm]{0mm}{5mm}OPE approximation& $f_D$ [MeV] & $f_{D'}$ [MeV]   \\ \hline\hline
\rule{0mm}{5mm} Perturbative LO& 120 & 123 \\
\rule{0mm}{5mm} Perturbative LO+NLO+NNLO  & 150 & 177 \\ \hline
\rule{0mm}{5mm} Perturbative  + $\langle \bar{q}q \rangle$(LO+NLO)  
& 190 & 142 \\
\rule{0mm}{5mm} Perturbative  + $\langle \bar{q}q \rangle$(LO+NLO)  +
$\langle d456\rangle $      & 194 & 137 \\
\hline\hline
\end{tabular}
\end{center}
\caption{\it Decay constants  of the $D$ and $D'$ mesons
from the sum rules with  different approximations for OPE  (procedure (I), central/default input). 
}
\label{tab:rel}
\end{table}
This brings us to the conclusion that the nonperturbative effects, most of all the leading
contribution of the quark condensate effectively redistribute the spectral density of the correlation function
so that the ground-state contribution gets enhanced and the first radial excitation suppressed. 

\section{Conclusion}

In this paper we have attempted to determine 
the decay constants of first radially excited
heavy-light mesons. To this end, modified 
QCD sum rules were obtained, in which, in addition to the ground-state meson contribution 
also the excited state is included being separated from the rest of the hadronic
spectral density. We applied two different procedures, one of them independent
of the ground-state contribution. The results of both procedures are consistent with each 
other and with the standard sum rule calculation \cite{GKPR} of the ground-state decay constants.
This consistency justifies the validity of the quark-hadron duality approximation 
beyond the ground state hadron in the correlation function of heavy-light currents.
In future, the same technique can be used for other excited hadrons with various 
flavour and spin-parity quantum numbers. Also studies of 
non-diagonal two-point correlation functions can be useful, with two different currents
having the same quantum numbers but different quark-gluon content.
Here, one will employ the conjecture that excited hadronic states have a larger coupling to quark-antiquark-gluon 
currents.  

Our results reveal a  relative suppression 
of the decay constants of the radially excited states with respect to the ground state.
This difference can be traced to nonperturbative effects.
In future, more precise data on radially excited  charmed and bottom mesons
will allow to improve the accuracy of our predictions.

The decay constant of the excited charmed meson $D'$  predicted here agrees
within uncertainties with  the recent lattice  QCD
result $f_{D'}=117\pm 25$  MeV obtained in ~\cite{Becirevic}. 
On the other hand, we cannot confirm the 
estimate   $f_{D^{*'}}\sim 300 $ MeV obtained in \cite{Ligeti}.

The results obtained in this paper can be used in several ways. 
First, it is possible to extend hadronic representations in  
various light-cone sum rules (LCSR) in order to try alternative 
patterns of quark-hadron duality, beyond one-resonance approximation.  
E.g., in the LCSR's for $B\to \pi$ form factors, in the channel of the $B$-meson
interpolating current, one can include in the hadronic spectral density
the term involving the first radial excitation $B'$ 
and the latter involves the decay constant $f_{B'}$. 
Furthermore, as already mentioned, one needs accurate information
on the excited charmed states including their  decay constants 
for the studies of $B\to D'$ form factors   \cite{Ligeti,Becirevic}.
A promising source of information on charmed resonances 
are nonleptonic decays of $B$ mesons  
to open-charmed  final states (see also \cite{Becirevic}).  
For three-body and four-body $B$ decays accurate Dalitz-plot analyses  
are among the primary goals of  current $B$-physics studies. 
These analyses demand reliable resonance-saturation models 
including the contributions of radially excited open charmed states. For the amplitudes
of these modes factorization estimates include the decay constants 
multiplied by heavy-to-light form factors.
To bring just one example,  the amplitude of 
$\bar{B}^0\to {\bar{D}^{*'-}_s} \pi^+$  decay contributes to the final 
state of three-body decay  $\bar{B}^0\to \bar{D}^0 K^- \pi^+$. 
The  factorizable part of $\bar{B}^0\to {{\bar{D}^{*'-}_s}} \pi^+$ is colour-enhanced, 
being proportional  to the product of the $B\to \pi$ form factor at 
the momentum transfer $q^2=m_{D^{*'}}^2$ and $f_{D^{*'}}$.
For the latter, the result of this study can immediately be used.

\section{Acknowledgments}
We thank Thorsten Feldmann for useful comments and 
Danny van Dyk for a
discussion on the statistical analysis.
This work is supported by  
DFG Research Unit FOR 1873 ``Quark Flavour Physics
and Effective Theories'',  Contract No.~KH 205/2-1. 

\end{document}